\documentclass[aps,10pt,notitlepage,nofootinbib,showkeys,prd]{revtex4-1}
\usepackage{amstext,amsmath,amssymb,amsfonts,bbm}
\usepackage[latin1]{inputenc}
\usepackage{epsfig}
\usepackage{hyperref}
\usepackage{amsthm}
\usepackage{subfigure}
\usepackage{color}
\usepackage{multirow}
\newcommand{\be}{\begin{equation}}
\newcommand{\ee}{\end{equation}}

\newcommand{\bea}{\begin{eqnarray}}	
\newcommand{\eea}{\end{eqnarray}}

\newcommand{\N}{\mathbbm{N}}

\DeclareMathOperator{\tr}{Tr}

\newtheorem{lemma}{Lemma}

\newtheorem{remark}{Remark}
\newtheorem{proposition}{Proposition}
\newtheorem{corollary}{Corollary}

\begin{document}

\title{\Large \bf Counting Line-Colored D-ary Trees}

\author{{\bf Valentin Bonzom}}\email{vbonzom@perimeterinstitute.ca}
\author{{\bf Razvan Gurau}}\email{rgurau@perimeterinstitute.ca}
\affiliation{Perimeter Institute for Theoretical Physics, 31 Caroline St. N, ON N2L 2Y5, Waterloo, Canada}

\date{\small\today}

\begin{abstract}
\noindent 
Random tensor models are generalizations of matrix models which also support a $1/N$ expansion. The dominant observables are in correspondence with some trees, namely rooted trees with vertices of degree at most $D$ and lines colored by a number $i$ from 1 to $D$ such that no two lines connecting a vertex to its descendants have the same color. 
In this Letter we study by independent methods a generating function for these observables. We prove that the number of such trees 
with exactly $p_i$ lines of color $i$ is 
$\frac{1}{\sum_{i=1}^D p_i +1} \binom{\sum_{i=1}^D p_i+1}{p_1} \dotsm \binom{\sum_{i=1}^D p_i+1}{p_D}$.
\end{abstract}

\medskip

%\noindent  Pacs numbers: 02.10.Ox, 04.60.Gw, 05.40-a

\maketitle

%%%%%%%%%%%%%%%%%%%%%%%%%%%%%
\section{Introduction}

The study of large random matrices in the past thirty years has successfully described measures which can be written as the exponential of 
single trace invariants perturbing a Gaussian. In addition to the standard Feynman diagrammatic expansion \cite{graph-combinatorics-difrancesco}, 
powerful methods exist to solve such models \cite{mm-review-difrancesco}, including orthogonal polynomials (which rely on eigenvalue decomposition) 
and more recently the topological expansion developed by Eynard \cite{topological-expansion}. The latter starts with the Schwinger-Dyson equations (also
 known as loop equations) written in terms of the resolvent $\tr \frac{1}{z-M M^\dagger}$ where $M$ is the random $N\times N$ matrix, and
 provides an intrinsic way to solve them at all orders in the $1/N$ expansion.

Random tensors are a generalization of random matrices to rank $D$ objects (having $D$ indices of size $N$ each). The study of their
large size statistics became possible thanks to the $1/N$ expansion discovered in \cite{largeN}. This expansion relies on the
construction of multi-unitary invariants, i.e. tensor contractions which are invariant under the external tensor product 
of $D$ copies of ${U}(N)$, each of them acting independently on each tensor index. In contrast with random matrices, there are many 
invariant monomials at a fixed degree. These monomials are indexed by colored graphs \cite{universality, uncoloring} (these colors correspond 
to the position of the index in a tensor contraction: in $T_{a_1 \dotsb a_D}$, $a_1$ has position, hence color, 1, up to $a_D$ which 
has color $D$). The $1/N$ expansion generalizes to any measure on a single random tensor that can be written as the exponential of 
such invariants \cite{virasoro}.

In the diagrammatic approach, the graphs contributing at leading order, known as \emph{melonic graphs} (equivalent to planar graphs of matrix 
models), have been identified in \cite{Bonzom:2011zz} enabling one to solve these models exactly at leading order \cite{uncoloring}. 
At the core of these solutions, a \emph{universality} theorem, first derived in \cite{universality}, states that all models are 
\emph{Gaussian} at large $N$ (but with a covariance which crucially depends on the joint distribution). In particular all invariant 
monomials corresponding to melonic graphs at fixed degree have the same expectation value. However, this certainly does not hold 
at sub-leading orders in the $1/N$ expansion, and different melonic graphs have different sub-leading corrections in $1/N$.

A method which would be fruitful to adapt to tensor models is the one developed by Eynard \cite{topological-expansion}. It first 
requires to introduce an equivalent of the resolvent. Writing the matrix resolvent 
like $\tr \frac1{z-M M^\dagger} =  \sum_{n\geq 0} z^{-n-1} \tr (M M^\dagger)^n$, it is tempting to generalize it by changing
 $\tr (M M^\dagger)^n$ with the sum over all invariants of degree $n$. As $z$ counts the degree of each invariant, this object
does not distinguish different invariants having the same degree. Although sufficient for 
the study of the leading order, this object should be refined in order to explore the finer structure of sub-leading orders,
for which a generating function which probes the structure of each invariant beyond their degree seems better adapted.

This Letter is a modest contribution in this direction. The dominant melonic invariants are in one-to-one correspondence 
with $D$-ary trees with lines colored $1$ up to $D$ such that no two lines connecting a vertex to its descendants 
have the same color. At leading order only the number of vertices matters, but at subsequent orders 
one must distinguish between different colored trees having the same number of vertices. As a first step, we study the generating 
function of these colored trees and find an explicit formula counting how many such trees with $p_i$ lines of color $i$ one can build. 
Countings of colored trees exist in the literature (like \cite{edge-colored}) but we could not find the
 counting which is relevant for our purpose.

In Section \ref{sec:stating} we introduce the problem and state our main results. The generating function is presented in Section \ref{sec:generating}
 and the proof of our results, based on a linear recursive sequence related to the generating function, is contained in Section \ref{sec:sequence}.

%%%%%%%%%%%%%%%%%%%%%%%%%%%%%
\section{Stating the problem} \label{sec:stating}

We consider a family of rooted trees defined by the following properties
\begin{itemize} 
 \item each vertex has at most $D$ descendants, $D\geq 2$,
 \item each line receives a color $i=1,\dotsc,D$ such that no two lines connecting a vertex to its descendants have the same color.
\end{itemize}
By adding leafs (univalent vertices) appropriately any such tree becomes a rooted $D$-ary tree with colored lines. 
Let $C_{p_1\dotsb p_D}$ be the number of such trees with exactly $p_i$ lines of color $i$ for all $i=1,\dotsc,D$. The purpose 
of this note is to derive an explicit formula for $C_{p_1\dotsb p_D}$.

Our strategy is based on the generating function 
\be \label{generating function}
F(g_1,\dotsc,g_D) = \sum_{p_1,\dotsc,p_D=0}^\infty C_{p_1\dotsb p_D}\ \prod_{i=1}^D g_i^{p_i}\;,
\ee
and the sequence $(F_n(g_i))_{n\in\N}$ defined by
\be \label{recursive sequence}
F_0 =1,\qquad F_n(g_1,\dotsc,g_D) = \sum_{p_1,\dotsc,p_D} C^n_{p_1 \dotsb p_D}\,\prod_{i=1}^D g_i^{p_i} \quad \text{with} \quad C^n_{p_1\dotsb p_D} = \frac{n}{\sum_{i=1}^D p_i +n}\,\prod_{j=1}^D \binom{\sum_{i=1}^D p_i +n}{p_j}
\;.
\ee
We will prove that $(F_n)$ is a linear recursive sequence whose characteristic equation is an algebraic equation satisfied by $F$. 
Examining the roots of this algebraic equation, we will obtain the following proposition.

\begin{proposition} \label{propF_n}
The sequence $(F_n(g_i))_{n\in\N}$ is a geometric sequence with common ratio $F$,
\be
F_n(g_1,\dotsc,g_D) = \Bigl(F(g_1,\dotsc,g_D)\Bigr)^n\;.
\ee
This implies that $F(g_1,\dotsc,g_D) = F_1(g_1,\dotsc,g_D)$, hence the following corollary.
\end{proposition}

\begin{corollary}
The number $C_{p_1\dotsb p_D}$ of rooted line-colored trees with maximal degree $D$ and exactly $p_i$ lines of color $i=1,\dotsc,D$ is
\be \label{prop}
C_{p_1\dotsb p_D} =C^1_{p_1\dotsb p_D}=  \frac{1}{\sum_{i=1}^D p_i +1}\,\prod_{j=1}^D \binom{\sum_{i=1}^D p_i +1}{p_j}\;.
\ee
\end{corollary}

%%%%%%%%%%%%%%%
\section{The generating function} \label{sec:generating}
%%%%%%%%%%%%%%%

The generating function \eqref{generating function} satisfies an algebraic equation which is obtained by simply observing that the root of a tree can have $k\leq D$ descendants connected by lines of colors $i_1,\dotsc,i_k$ all different. Therefore
\be \label{tree recursion}
\begin{aligned}
F(g_1,\dotsc,g_D) &= 1 + \bigl(\sum_{1\le i_1\le D } g_{i_1} \bigr) F + \bigl(\sum_{1\le i_1<i_2 \le D} g_{i_2} g_{i_2}\bigr) F^2 + \dotsb 
   + \bigl( g_1 \dotsb g_D \bigr) F^D\;,\\
&= \sum_{k=0}^D \biggl( \sum_{1\leq i_1<\dotsb <i_k\leq D}\, \prod_{l=1}^k g_{i_l} \biggr)\ [F(g_1,\dotsc,g_D)]^k \;,
\end{aligned}
\ee
and $F$ is the root $x_{(0)}(g_1,\dots g_D)$ of this polynomial equation such that
\bea
\lim_{g_1,\dots g_D \to 0} x_{(0)} (g_1,\dots g_D)=1 \; .
\eea

The following lemma show how to distinguish the desired root $x_{(0)}(g_1,\dots g_D) $ from the other roots of the above polynomial.
\begin{lemma} \label{lem:x_0}
 For all $R>1$, there exists $\epsilon_R>0$ such that for all $|g_1|, \dotsc, |g_D|<  \epsilon_R$, the polynomial 
 \be \label{eq:charD}
 Q(X) = - X + \sum_{k=0}^D \Bigl( \sum_{1\leq i_1<\dotsb <i_k\leq D}\, \prod_{l=1}^k g_{i_l} \Bigr) X^k
 \ee
 has exactly one root $x_{(0)}$ with $|x_{(0)}|<R$, all other roots $x_{(i)}$ for $i\neq 0$ having norms $|x_{(i)}|\ge R$. In particular $x_{(0)}=F$ as it is the only root satisfying $ \lim_{g_i\to 0} x_{(0)} =1$.
\end{lemma}

{\bf Proof.} To establish this lemma we use Rouch\'e's theorem whose statement is now recalled. Let $f, g$ be two holomorphic functions and $S$ a closed contour which does not contain zeros of $f$ and $g$. If for all $z\in S$
\be
|f(z) - g(z)|< |g(z)|\;,
\ee
then the number of zeros of $f$ and the number of zeros of $g$ encircled by $S$ (counted with multiplicities) are the same.

We now exploit this theorem to get bounds on the absolute value of the roots of the polynomial \eqref{eq:charD}. Take
\be
f(z) =  Q(z)\;,\qquad \text{and} \qquad g(z) = -z \; .
\ee
Let $R>1$, and $S = \{ z\in\mathbbm{C}, |z|=R\}$ be the circle of radius $R$. Note that on $S$, $|g(z)|=R$. Set $\epsilon_R>0$ with $\epsilon_R<\frac{R^{1/D}-1}{R}$ so that on $S$,
\be
|f(z) - g(z)| \leq \sum_{k=0}^D \Bigl( \sum_{1\le i_1<i_2\dots <i_k\le D} |g_{i_1}| \dotsm |g_{i_k}| \Bigr) |z|^k
 < \sum_{k=0}^D \binom{D}{k} (\epsilon_R\, R)^k
 \leq (1+\epsilon_R R)^D
 \leq |g(z)| \; .
\ee
As $g(z)$ has an unique root $z=0$ inside the circle of radius $R$, we conclude that $f(z)$ also has exactly one root,
which we denote $x_{(0)}$, with $ |x_{(0)}| <R $. As this is the only root which can go to one when $g_i \to 0$, it is identified with the generating function $F$.
\qed

\begin{remark}
 At $D=2$, $Q(X)= 1 +(g_1+g_2-1)X + g_1 g_2 X^2$ and $Q(x)=0$ is easily solved,
\be
x_{(0)} = \frac{1-g_1-g_2-\sqrt{(1-g_1-g_2)^2-4g_1 g_2}}{2\,g_1 g_2}\qquad \text{and} \qquad x_{(1)} = \frac{1-g_1-g_2+\sqrt{(1-g_1-g_2)^2-4g_1 g_2}}{2\,g_1 g_2}\;.
\ee
\end{remark}

%We will show that $(F_n)$ satisfies a linear recursion whose characteristic equation is \eqref{tree recursion}. We will finally conclude using a lemma which distinguishes $x_{(0)}$ from the other root of \eqref{tree recursion}.

%%%%%%%%%%%%%%%%%%%%%%%%%%
\section{The linear recursive sequence} \label{sec:sequence}
%%%%%%%%%%%%%%%%%%%%%%%%%%

%%%%%%%%%%
%\subsection{}
%%%%%%%%%%

First we show that the sums defining each $F_n(g_i)$ in \eqref{recursive sequence} converge absolutely when $|g_i|< \frac{(D-1)^{D-1}}{D^D}$ for all $i=1,\dotsc,D$. Let $\epsilon>0$ such that $\epsilon<\frac{(D-1)^{D-1}}{D^D}$, then for all complex $g_1,\dotsc,g_D$ with norm $|g_i|<\epsilon$
\be
|F_n(g_1,\dotsc,g_D)| \leq \sum_{p=0}^{\infty} \epsilon^p \frac{n}{p+n} \sum_{\substack{ \{p_i\}\\ \sum_i p_i = p}} 
 \binom{p+n}{p_1} \dotsm  \binom{p+n}{p_D}
\ee
The sums over $p_i$ are computed by equating the coefficients of $x^p$ in $(1+x)^{Dp+Dn}$ and $(1+x)^{p+n} \dotsm (1+x)^{p+n}$, hence
\bea\label{eq:Dcatalan}
|F_n(g_1,\dotsc,g_D)| \leq \sum_{p=0}^{\infty}  \frac{n}{p+n} \binom{Dp+Dn}{p} \; \epsilon^p \; .
\eea
One finds using the Stirling formula that the radius of convergence of the above series is $\frac{(D-1)^{D-1}}{D^D}> \epsilon$, hence $F_{n}(g_1,\dotsc,g_D)$ converges absolutely.

%%%%%%%%%%
%\subsection{}
%%%%%%%%%%

\begin{lemma}\label{lem:recD}
The sequence $(F_n)$ respects the recursion
\be
\forall n\ge 0 \qquad 
F_{n+1}  =  F_n + \sum_{k=1}^D \biggl( \sum_{1\leq i_1<\dotsb <i_k\leq D} \prod_{l=1}^k g_{i_l} \biggr)\ F_{n+k} \;.
\ee
\end{lemma}
{\bf Proof:} The recursion translates into
\bea\label{eq:recD}
 \forall p_i \ge 1\qquad
 C^{n+1}_{p_1,\dots p_D} = C^n_{p_1,\dots p_D} + \sum_{k=1}^{D}  \sum_{1\le i_1<i_2\dots <i_k\le D}
 C^{n+k}_{p_1,\dots p_{i_1}-1,\dots p_{i_k}-1,\dots p_D}  \; .
\eea
The boundary cases, when some $p_i=0$, just reproduce the recursion at level $D-1$. Let us denote $P=\sum_{i=1}^D p_i$. The right hand side of \eqref{eq:recD} is  
\be
\frac{(P+n)^{-1} \bigl[(P +n)! \bigr]^{D} }{ \prod_{i=1}^D p_i!(P -p_i +n +1)!}
\Bigl[ n \prod_{i=1}^D (P -p_i +n +1)  + \sum_{k=1}^D (n+k) \sum_{ 1\leq i_1<\dotsb <i_k\leq D  } \prod_{l=1}^k p_{i_l} \prod_{j\neq i_1,\dotsc,i_k}  (P -p_j +n +1) \Bigr] \; .
\ee
We write $n \prod_{i=1}^D (P -p_i +n +1) = (n+1)\prod_{i=1}^D (P -p_i +n +1) - \prod_{i=1}^D (P -p_i +n +1)$ so as to re-arrange the square bracket above as
\bea\label{eq:lung}
&& n \prod_i (P -p_i +n +1)  + \sum_{k=1}^D (n+k) \sum_{ 1\leq i_1<\dotsb <i_k\leq D  } \prod_{l=1}^k p_{i_l} \prod_{j\neq i_1,\dotsc,i_k}  (P -p_j +n +1)  \crcr
&&
= (n+1) \prod_{i=1}^D ( P +n +1) -  \prod_{i=1}^D (P -p_i +n +1) +  \sum_{k=1}^D (k-1) 
\sum_{ 1\leq i_1<\dotsb <i_k\leq D  } \prod_{l=1}^k p_{i_l} \prod_{j\neq i_1,\dotsc, i_k} (P -p_j +n +1) \crcr
&& = (n+1) (P +n) (P +n +1)^{D-1} + (n+1)  (P +n +1)^{D-1} - \prod_{i=1}^D (P -p_i +n +1) \crcr
&& +  \sum_{ k=1  }^D (k-1) 
\sum_{ 1\leq i_1<\dotsb <i_k\leq D} \prod_{l=1}^k p_{i_l} \prod_{j\neq i_1,\dots i_k} 
(P-p_j +n +1) \; .
\eea
The first term of the last equality is exactly what is needed to form $C^{n+1}_{p_1\dotsb p_D}$. Therefore we focus now on the sum of the three other contributions,
\bea
&& (n+1) (P+n+1)^{D-1} - \prod_{i=1}^D (P+n+1-p_i) + \sum_{k=2}^D (k-1) 
\sum_{ 1\leq i_1<\dotsb <i_k\leq D} \prod_{l=1}^k p_{i_l} \prod_{j\neq i_1,\dots i_k} (P +n +1  -p_j  ) \crcr
&& = (n+1) (P+n+1)^{D-1} - (P+n+1)^D + (P+n+1)^{D-1} P  + \sum_{k=2}^D(-)^{k+1} \sum_{ 1\leq i_1<\dotsb <i_k\leq D} \Bigl[\prod_{l=1}^k p_{i_l}\Bigr] (P+n+1)^{D-k} \crcr
&& + \sum_{ k=2  }^D (k-1) \sum_{ 1\leq i_1<\dotsb <i_k\leq D} \prod_{l=1}^k p_{i_l} 
\sum_{m=0}^{D-k} ( P+n+1 )^{D-k-m} (-)^m \sum_{\substack{1\leq j_1<\dotsb <j_m \leq D\\j_t\neq i_l}} \prod_{t=1}^m p_{j_t} \; .
\eea
The first three terms cancel. For the remaining terms
\begin{multline}\label{eq:sum-sum}
\sum_{k=2}^D(-)^{k+1} \sum_{ 1\leq i_1<\dotsb <i_k\leq D} \Bigl[\prod_{l=1}^k p_{i_l}\Bigr] (P+n+1)^{D-k} \\
 + \sum_{ k=2  }^D (k-1) \sum_{ 1\leq i_1<\dotsb <i_k\leq D} \prod_{l=1}^k p_{i_l} 
\sum_{m=0}^{D-k} ( P+n+1 )^{D-k-m} (-)^m \sum_{\substack{1\leq j_1<\dotsb <j_m \leq D\\j_t\neq i_l}} \prod_{t=1}^m p_{j_t}\;,
\end{multline}
we take into account that $k+m$ ordered integers can be partitioned into $\binom{k+m}{k}$ ways into two subsets
of $k$ and $m$ ordered integers. Thus the second sum rewrites as a sum over $q=k+m$ 
\bea
 \sum_{q=2}^D\ \sum_{ 1\leq i_1<\dotsb <i_q\leq D  } p_{i_1} \dotsm p_{i_q} (-)^q (P+n+1)^{D-q} 
 \sum_{k=2}^q (-)^k (k-1) \binom{q}{k} \; .
\eea
But 
\be
\sum_{k=2}^q (-)^k (k-1) \binom{q}{k} = 1 + \sum_{k=0}^q (-)^k (k-1) \binom{q}{k} = 1 - (1-1)^q - q(1-1)^{q-1} =1 \;,
\ee
hence the whole quantity displayed in \eqref{eq:sum-sum} is zero and the lemma \ref{lem:recD} follows.
\qed

\medskip

Therefore, $F_{n+D}$ is obtained recursively from the set $(F_{p})_{p<n+D}$. The characteristic polynomial of this recursion is exactly $Q(X)$ (Equation \eqref{eq:charD}). It means that the solution $x_{(0)}$ is one of the common ratios of $(F_n)$. We denote the others $x_{(j)}$, and assuming they are all different,
\be \label{formal F_n}
 F_n = a\, x^n_{(0)} + \sum_j b_j\, x^n_{(j)} \;,
\ee
for some functions $a(g_1,\dotsc,g_D), b_j(g_1,\dotsc,g_D)$ which can in principle be determined by $D$ initial conditions. However, we cannot use the initial conditions (remember we want to prove that $F=F_1$) so we have to proceed differently. Each common ratio in the sum \eqref{formal F_n} is controlled thanks to the Lemma \ref{lem:x_0}, as $x_{(0)}$ is bounded from above and each $x_{(j)}$ is bounded from below. Now we need to control the sequence $(F_n)$ independently of its common ratios. This is done through the following lemma.

\begin{lemma} \label{lem:boundD}
 For all $K>1$, there exists  $\epsilon_K >0$ such that for all $g_1,\dotsc, g_D \in \mathbb{C}$ with $|g_1|,\dotsc,|g_D| < \epsilon_K $,  $F_n(g_1,\dots g_D)$ is polynomially bounded by $K$,
\be
 \forall n \ge 0\qquad |F_n(g_1,\dots g_D)| \leq K^n \;.
\ee
\end{lemma}

{\bf Proof.} For $n=0$, this is trivial as $F_0=1$. Let thus be $n\geq1$ and $K>1$. It is enough to chose $\epsilon_K$ such that 
\bea
\epsilon_K < \frac{(D-1)^{D-1}}{D^D} \; , \qquad  \epsilon_K < \frac{1}{2De}
 \qquad \text{and} \qquad e^{2D\epsilon_K}+ \frac{1}{\sqrt{2\pi} } \frac{2De\epsilon_K}{1-2De\epsilon_K} <K \; .
\eea
With $\epsilon_K < \frac{(D-1)^{D-1}}{D^D}$, one can use Equation \eqref{eq:Dcatalan} which implies
\bea
 |F_n| \leq \sum_{p=0}^{\infty} \frac{n}{n+p} \frac{(Dn+Dp)^p}{p!} \; \epsilon_K^p
\leq \sum_{p=0}^{\infty} \frac{ (n+p)^{p} }{p!} (D\epsilon_K)^p \; . \crcr
\eea
We use the fact that $(n+p)^{p} \leq (2n)^{p}$ when $p\leq n$ and $(n+p)^{p} \leq (2p)^{p}$ when $p\geq n$ to obtain the bound
\bea
&&  |F_n| \leq  \sum_{p=0}^{n} n^p \frac{(2D\epsilon_K)^p}{p!} + \sum_{p=n}^{\infty} \frac{p^{p}}{p!}(2D\epsilon_K)^p 
\leq    e^{2Dn\epsilon_K} +   \sum_{p=n}^{\infty} \frac{p^{p}}{p!} (2D\epsilon_K)^p \;.
\eea
Now we use $p! \geq \sqrt{2\pi p} \; e^{p\ln p -p}, \forall p\geq1$ and as $\epsilon_K < \frac{1}{2De}$ we get
\be
\begin{aligned}
 |F_n| &\leq e^{2Dn\epsilon_K} + \frac{1}{\sqrt{2\pi} } \sum_{p=1}^{\infty} \frac{1}{\sqrt{p}} \; (2De\epsilon_K)^p
 \leq   e^{2Dn\epsilon_K} +  \frac{1}{\sqrt{2\pi} } \sum_{p=1}^{\infty} (2De\epsilon_K)^p \\
 &\leq e^{2Dn\epsilon_K} +  \frac{1}{\sqrt{2\pi} } \frac{2De\epsilon_K}{1-2De\epsilon_K} \leq
 \Bigl( e^{2D\epsilon_K}+ \frac{1}{\sqrt{2\pi} }  \frac{2De\epsilon_K}{1-2De\epsilon_K} \Bigr)^n \; .
\end{aligned}
\ee
\qed

We are now in position to prove Proposition \ref{propF_n}, by combining Lemmas \ref{lem:x_0} and \ref{lem:boundD}. Choose $R>K>1$ and consider $|g_i|<\inf (\epsilon_R,\epsilon_K)$ with $\epsilon_R, \epsilon_K$ as in the lemmas. First we consider the case where all $x_{(j)}$ have different norms and denote $x_{(\rm max)}\neq 0$ the one with the largest norm. In particular $|x_{(\rm max)}| \geq R>1$. At large $n$, the norm of $F_n$ is dominated by $b_{({\rm max})} x_{(\rm max)}$. Hence there exist a constant $A>0$ and an integer $N$ such that for all $n\geq N$,
\be
|F_n(g_1,\dotsc,g_D)|\geq A\,|b_{({\rm max})}|\,|x_{\rm max}|^n \geq A\,|b_{({\rm max})}|\,R^n\;.
\ee
However $|F_n|\leq K^n$ with $R>K$. Therefore we conclude $b_{({\rm max})}=0$. We can repeat this reasoning with the root $x_{(j)}$ that has the second largest norm, and so on until we get $F_n = a x_{(0)}^n$. The initial condition $F_0=1$ for all $g_i$ finally leads to $F_n = x_{(0)}^n$.

The case where some of the roots have the same norm is quite similar. The idea is to extract sub-sequences $(F_{r(n)})$ for which $F_{r(n)}$ behaves at large $n$ like a coefficient times some combination of the roots $x_{(j)}$, $j\neq 0$, where this combination is greater than $R^n$ when $|x_{(j)}|\geq R$.

\begin{remark}
 At $D=2$, the number of line-colored trees with $p_1$ lines of color 1 and $p_2$ lines of color 2 is $C_{p_1 p_2} = \frac{1}{p_1+p_2+1} \binom{p_1+p_2+1}{p_1} \binom{p_1+p_2+1}{p_2}$. These numbers are known as the Narayana numbers $N(p_1+p_2+1,p_1+1)$.
\end{remark}

\begin{remark}
 By summing the numbers $C_{p_1 \dotsb p_D}$ over all possible numbers of lines of each color at a fixed total number of lines $P -1$,
 \be
 \sum_{\substack{ \{p_i\} \\ \sum_i p_i =P-1}} C_{p_1\dotsb p_D} = \frac{1}{P}\,\binom{D(P-1) +D}{P-1} = \frac{1}{DP+1}\,\binom{DP+1}{P}\;,
 \ee
 we obtain the total number of $D$-ary trees on $P$ vertices also known as the $D$-Catalan numbers (pp. 200 in \cite{graham-knuth-patashnik}, proposition 6.2.2 in \cite{stanley} and more details in \cite{Bonzom:2011zz}).
\end{remark}

\begin{remark}
 Proposition \ref{propF_n} implies that $F_n F_m = F_{n+m}$, corresponding to interesting combinatorial identities,
 \be
 \sum_{\{k_i=0,\dotsc,p_i\}} C^{n}_{k_1\dotsb k_D}\,C^m_{p_1-k_1 \dotsb p_D-k_D} = C^{n+m}_{p_1\dotsb p_D}\;.
 \ee
 For example, when $D=2$ one gets
 \begin{multline}
 \sum_{k_1,k_2=0}^{p_1,p_2} \frac{n\ m}{(k_1+k_2+n)\,(p_1-k_1+p_2-k_2+m)} \binom{k_1+k_2+n}{k_1} \binom{k_1+k_2+n}{k_2} \\
 \times \binom{p_1-k_1+p_2-k_2+m}{p_1-k_1} \binom{p_1-k_1+p_2-k_2+m}{p_2-k_2}
 = \frac{n+m}{p_1+p_2+n+m} \binom{p_1+p_2+m+n}{p_1} \binom{p_1+p_2+m+n}{p_2}\;.
 \end{multline}
\end{remark}

%%%%%%%%%%
%\subsection{Conclusion}
%%%%%%%%%%

%%%%%%%%%%%%%%%%%%%%%%%%%%%%%%
\section*{Acknowledgements}

Research at Perimeter Institute is supported by the Government of Canada through Industry Canada and by the Province of Ontario through the Ministry of Research and Innovation.

%%%%%%%%%%%%%%%%%%%%%%%%%%%

\end{document}